\documentclass[
aip,reprint]{revtex4-2}
\usepackage[normalem]{ulem}
\usepackage{graphicx}
\usepackage{dcolumn}

\usepackage[utf8]{inputenc}
\usepackage[T1]{fontenc}
\usepackage{mathptmx}
\usepackage{hyperref}

\usepackage{color}
\usepackage[textsize=tiny]{todonotes}

\usepackage{xspace}

\newcommand{\perroothz}{$/\sqrt{\textrm{Hz}}$\xspace}

\begin{document}
\title{Limits to the sensitivity of a rare-earth-enabled cryogenic vibration sensor}
\date{\today }
\author{Anne Louchet-Chauvet}
\affiliation{ESPCI Paris, Universit\'e PSL, CNRS, Institut Langevin, 75005 Paris, France}
\author{Thierry Chaneli\`ere}
\affiliation{Univ. Grenoble Alpes, CNRS, Grenoble INP, Institut N\'eel, 38000 Grenoble, France}

\date{\today}

\begin{abstract}
Cryogenics is a pivotal aspect in the development of quantum technologies. Closed-cycle devices have recently emerged as an environmentally friendly and low-maintenance alternative to liquid helium cryostats. Yet the larger level of vibrations in dry cryocoolers forbids their use in most sensitive applications. 
In a recent work, we have proposed an inertial, broadband, contactless sensor based on the piezospectroscopic effect, \emph{ie} the natural sensitivity of optical lines to strain exhibited by impurities in solids. This sensor builds on the exceptional spectroscopic properties of rare earth ions and operates below $4$~K, where spectral hole burning considerably enhances the sensitivity.
In this paper, we investigate the fundamental and technical limitations of this vibration sensor by comparing a rigid sample attachment to cold stage of a pulse-tube cryocooler and a custom-designed exchange gas chamber for acoustic isolation.
\end{abstract}


\maketitle


\section{Introduction}

The recent global effervescence around quantum technologies has led to an increasing need for reliable and efficient cryogenic systems. Continuously-operated closed-cycle cryocoolers are progressively replacing liquid helium cryostats that suffer from heavy logistics and high costs \cite{kramer2020helium}. However the low running cost and ease of use of dry systems comes at the price of a high level of acoustic vibrations generated by the cycling gas flow. These vibrations are problematic for many cryogenic experiments, including cavity QED~\cite{vadia2021open_final}, trapped ions spectroscopy~\cite{micke2019closed}, quantum memories~\cite{gundogan2015solid} and frequency references based on rare-earth ions in crystals~\cite{thorpe2013shifts,gobron2017dispersive}, scanning probe microscopy~\cite{quacquarelli2015scanning}, but also in the field of astronomy with bolometers~\cite{maisonobe2018vibration} or gravitational waves detectors~\cite{tomaru2004vibration}.

Relevant diagnosis of the vibrations in a cryostat is a necessity to ensure correct operation of the experiment. The specifications provided by commercial suppliers often prove insufficient to assess the actual impact of vibrations on a given experiment, either because the actual setup mechanical assembly differs from the nominal configuration, or because the information is incomplete (often limited to low acoustic frequencies of the order or below $1$~kHz and/or peak-to-peak values).

In a recent publication~\cite{louchet2019piezospectroscopic} we have proposed an original optical inertial vibration sensor design with a high-frequency range (up to $1$~MHz). This sensor relies on the piezospectroscopic effect in rare-earth doped solids. In this work we investigate the sensitivity of this method. To that end, we explore various vibrational environments with the help of a customized pulse-tube cryocooler.

\section{Measuring vibrations with a rare-earth doped crystal}
\label{sec:measvib}
Due to the screening of the outer electronic shells, the $4f$ energy levels of a rare-earth ion embedded in a crystal exhibit narrow transitions whose position is mainly determined by the crystal field created by the surrounding ligands. Modifying the interatomic distance, e.g. in a static way by applying high pressure, is a convenient way to explore the crystal structure and the site symmetry~\cite{bungenstock2000effect, kaminska2016spectroscopic}. This sensitivity to strain, also known as the piezospectroscopic effect~\cite{kaplyanskii1964noncubic}, translates into a sensitivity to vibrations that is generally considered as an obstacle to high resolution measurements in rare-earth-based systems, especially in closed-cycle cryocoolers~\cite{tomaru2004vibration}. For this reason, custom solutions have been developed to reduce the transmission of vibrations in such devices while still ensuring a good thermal contact~\cite{thorpe2013shifts,chen2016coupling,gobron2017dispersive}.

We recently proposed to take advantage of the sensitivity of rare-earth-doped crystals to vibrations to provide a local measurement of the mechanical stability of a sample holder in a cryocooler~\cite{louchet2019piezospectroscopic}. The proposed method consists in attaching a rare-earth ion-doped crystal to the holder under test and measuring in real time the transmission of a laser beam tuned to the center of a narrow spectral hole whose linewidth can be adjusted from a few tens of kHz to a few MHz.

The piezospectroscopic effect in a solid is generally described with a tensor to account for its anisotropic nature. However, in most rare-earth ion-doped crystals, the optically active ions occupy several orientationally inequivalent crystallographic sites~\cite{sun2000symmetry}. In addition, the vibration-induced strain is in practice non-uniform within the crystal~\cite{zhang2020inhomogeneous}. Therefore, instead of causing a time-dependent shift of the spectral hole frequency, the vibrations result in a time-dependent \emph{broadening} effect on the spectral hole, characterized by a scalar sensitivity $\kappa$ (expressed in Hz/Pa).
Using a simple toy model describing the propagation of acoustic waves within a mm-sized crystal and assuming a conservation of the spectral hole area, the absorption at the center of the hole $\alpha(t)$ is linked to the instantaneous atomic velocity of the rare-earth ions~\cite{louchet2019piezospectroscopic}:
\begin{equation}
\kappa \frac{E}{\mathcal{V}}\left|\dot{x}(t)\right|
=\frac{\alpha (t)}{\alpha_0-\alpha(t)}\Gamma_{\mathrm{HB}}
\label{eq:abs}
\end{equation}
where  $E$ is the Young modulus of YAG and $\mathcal{V}$ is the sound velocity in YAG. $\alpha(t)$ is the time-dependent crystal absorption at the center of the spectral hole and $\alpha_0$ is the spectral hole depth. $\Gamma_{\mathrm{HB}}$, the width of the spectral hole, is periodically measured by chirping the laser around the central frequency. Due to the transient nature of the spectral hole, this equation is only valid for acoustic frequencies above $1/T_1$, where $T_1$ is the spectral hole lifetime ($10$~ms in our case). We point out that while our model only considers monodirectional vibrations, the multiple crystallographic site orientations and the Poisson effect (characterized by the Poisson's ratio of the crystal) provide sensitivity to vibrations in all directions.

\section{Vibration measurement sensitivity}

\subsection{Optical setup}
The laser is a sub-kHz linewidth extended cavity diode laser (ECDL)~\cite{crozatier2004laser}, tuned to the center of the Tm:YAG $^3H_6\rightarrow ^3H_4$ line ($793.37~$nm). It is installed on a separate optical table to avoid coupling to the cryocooler vibrations. The cryostat is equipped with optical windows so we focus the laser beam in the crystal ($100\mu$m waist). We collect the transmitted part that is measured with an avalanche photodetector (Thorlabs APD110A). For more detail on the experimental setup, the reader is referred to~\cite{louchet2019piezospectroscopic}.

\subsection{Cryocooler design}
We use a modified pulse-tube cryocooler TransMIT PTD-009 with an Oerlikon COOLPAK 2000A compressor. The modifications brought to the cryostat consist in an inner gas chamber (IGC) made of OFHC copper, attached to the cold finger. This chamber is equipped with two windows. It may optionally be filled with gas via a thin tube connected to a room temperature helium gas tank. The stainless steel injection tube is thermalized at the first stage of the cryocooler using a copper clamp and braids. Optimized design of the tube thickness (0.1mm for a 3.2mm diameter) and length (1.5m) ensures a negligible thermal load on both first and second stages.
The crystal is resting on an OFHC sample holder placed inside the IGC (see also Figure~\ref{fig:drum} for schematic 3D-visualisation). The thermal and mechanical contact is achieved using a thin layer of Apiezon-N grease that hardens at cryogenic temperatures~\cite{kreitman1972correlation}. This hardening ensures an efficient transmission of the vibrations through the grease layer, but may lead to some degree of static stress at the contact surface, which results in a marginal additional contribution to the static inhomogeneous broadening of the optical line. This has no effect on the spectral hole width. Even when the crystal itself can be thermally cooled by the exchange gas that may fill the IGC, its small surface (a few tens of mm$^2$) strongly limits direct cooling of the sample. The suspended sample holder with a much larger surface (a few tens of cm$^2$) is contrariwise efficiently cooled by the exchange gas.

\begin{figure}[t]
\centering
\includegraphics[width=6cm]{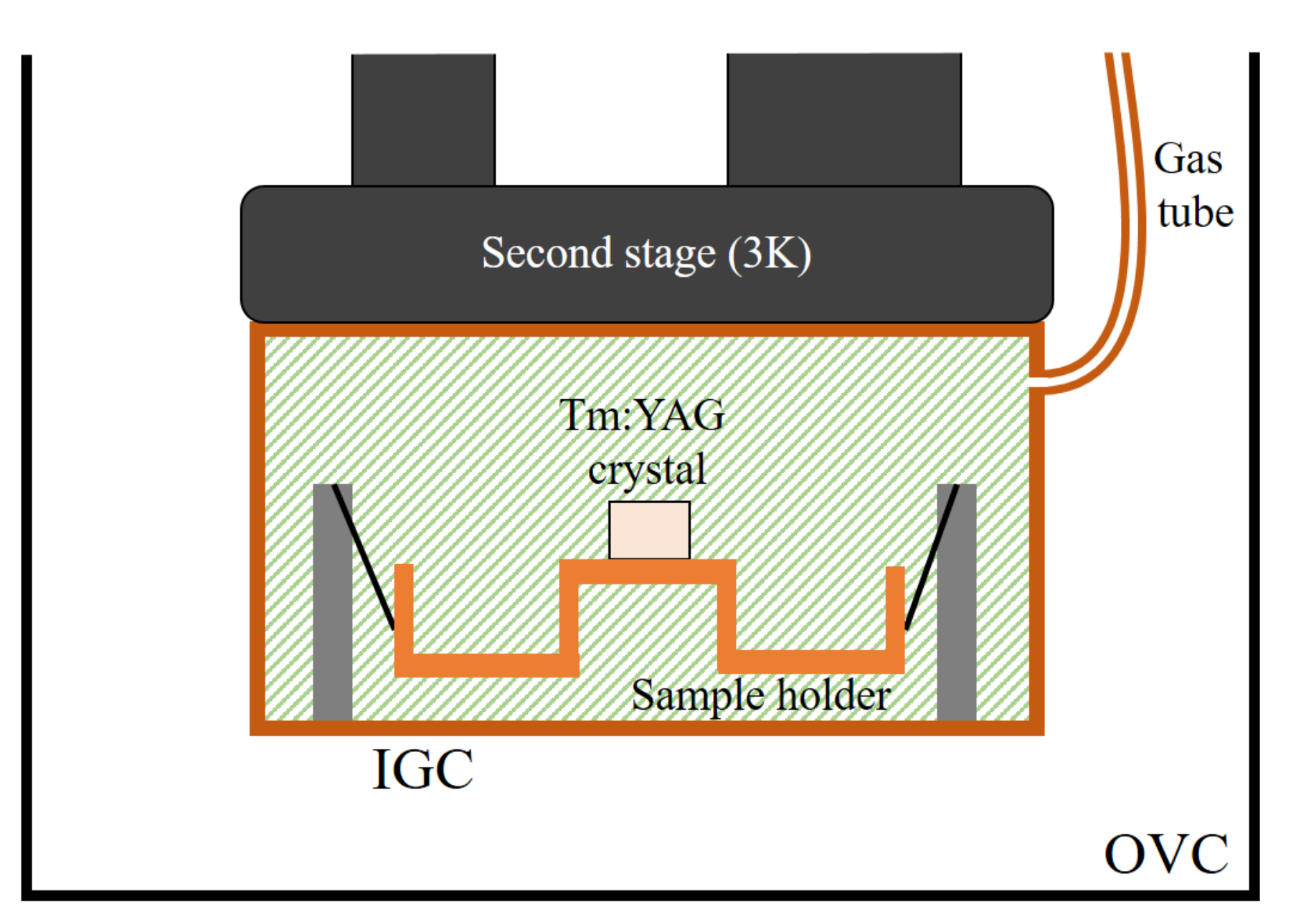}
\caption{Schematic cryostat cold part in configurations (b) and (c): the holder is suspended to the posts and the IGC is filled with helium gas. In configuration (a) the sample holder rests at the bottom of the IGC. (OVC: outer vacuum chamber. IGC: inner gas chamber.). The IGC and OVC are equipped with optical windows (not shown). A copper radiation shield (wrapped with multi-layer insulation), not represented for simplicity, is attached to the first stage.}
\label{fig:IGC}
\end{figure}

Three experimental configurations are investigated in order to explore different vibrational environments for our sensor.
In configuration (a), the sample holder is resting at the bottom of the IGC, with a thin layer of thermal grease to ensure good thermal contact and rigid mechanical coupling. The IGC and the tube are under vacuum. In this configuration the pulse tube vibrations are expected to couple efficiently to the crystal.
In configurations (b) and (c), the sample holder is suspended by nylon threads to three stainless steel posts attached to the bottom of the IGC (see Figure~\ref{fig:IGC}). Such a suspension system is expected to significantly damp the vibrations transmitted to the sample, similarly to the vibration isolation techniques developed for gravitational detectors~\cite{ushiba2021cryogenic}.
Helium gas is admitted through the tube to reach a pressure around $100$~mbar, ensuring thermalization of the sample holder on which the crystal is contacted. In configuration (b), the compressor is on, generating mechanical vibrations that are attenuated through the suspension threads and the gas. In configuration (c), the compressor is switched off for just a few seconds so that the sample temperature does not rise by more than $0.5$~K. The three configurations are summarized in Table~\ref{table:exp}.

\begin{table}
\begin{tabular}{|c|c|c|}
\hline
$\#$ & mechanical contact & compressor\\
\hline
(a) & rigid &  on\\
(b) & suspension in gas & on \\
(c) & suspension in gas & off \\
\hline
\end{tabular}
\caption{Experimental configurations allowing the exploration of three vibrational environments in the cryocooler.}
\label{table:exp}
\end{table}

We additionally monitor the acoustic environment up to $20$~kHz with the help of a commercial soundmeter (RadioShack 33-099) attached to the outer cold head case of the cryostat. This device records the acoustic environment in the laboratory.

\subsection{Spectral hole shapes}

\begin{figure}[t]
\centering
\includegraphics[width=8.5cm]{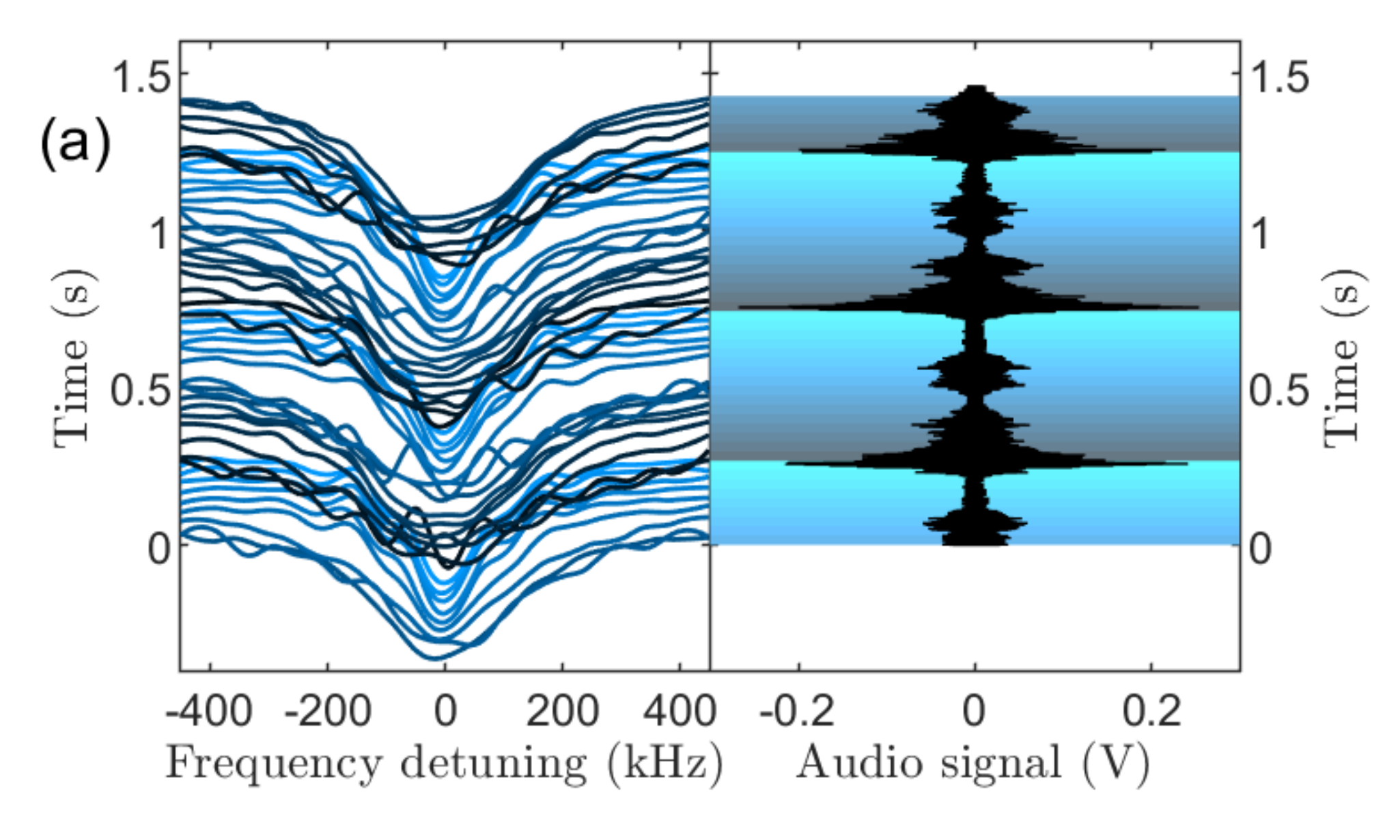}
\includegraphics[width=8.5cm]{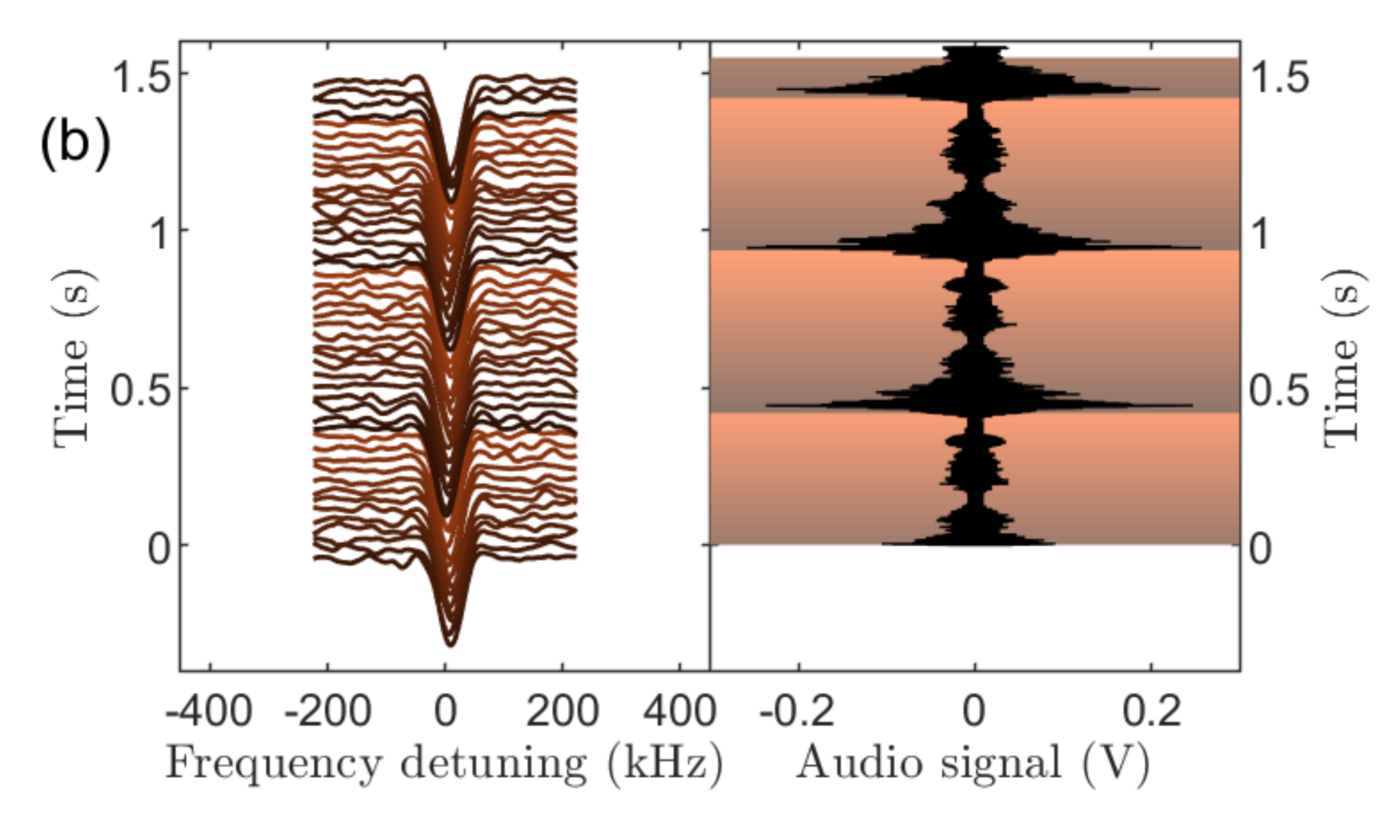}
\includegraphics[width=8.5cm]{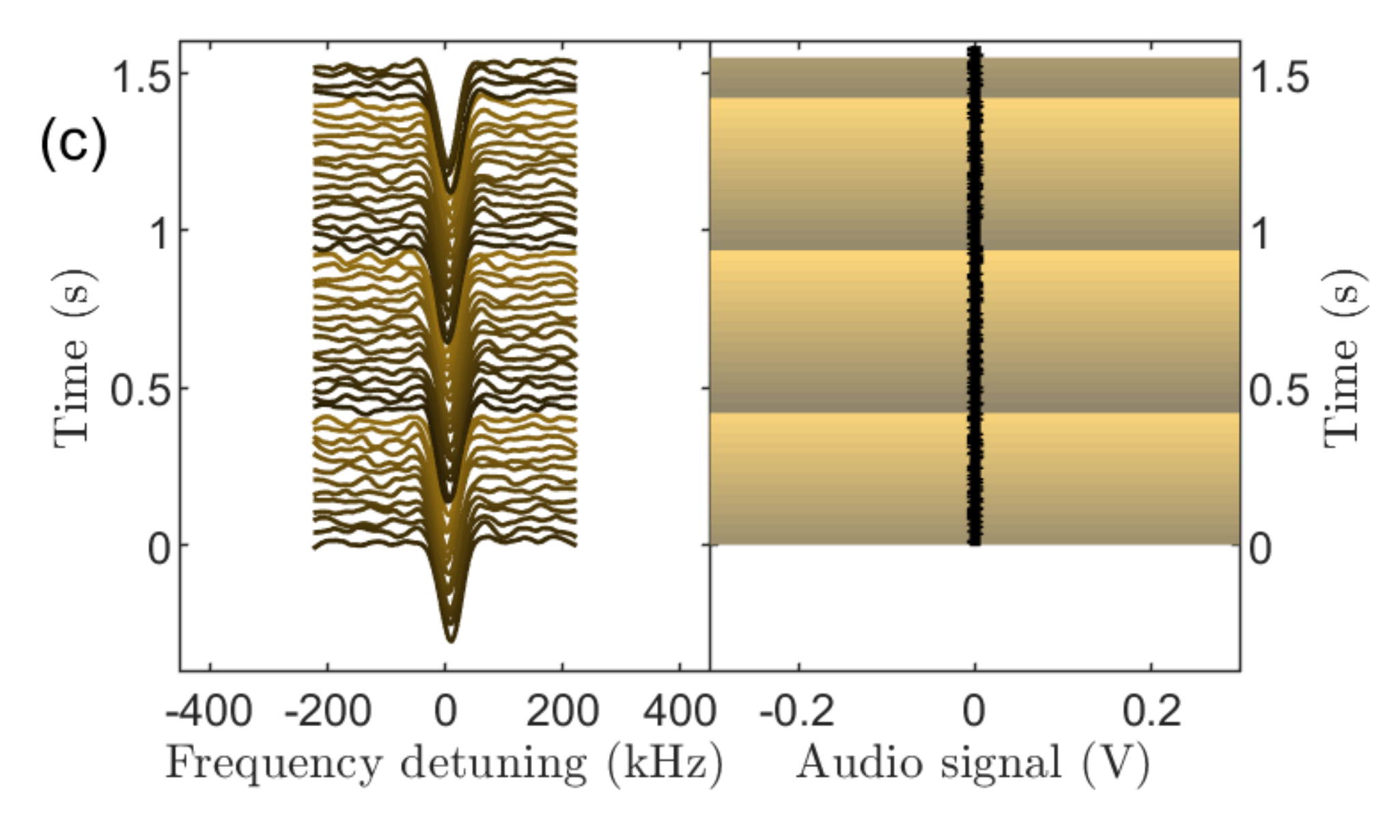}
\caption{Left: Spectral hole absorption throughout a repetition of 3 cycles (0.5s duration) of the rotary valve in configuration (a), (b) and (c) (see Table~\ref{table:exp}). The graphs are vertically offset for a clear view of the spectral hole shape evolution. Right: Spectral hole width (red triangles) and audio signal simultaneously recorded (black line). Each moment in the valve cycle is identified with a color gradient (from dark at the beginning to bright at the end).}
\label{fig:trous}
\end{figure}

We first examine the spectral hole profile evolution in the three cryostat configurations, displayed in Figure~\ref{fig:trous}, together with the audio signal recorded with the soundmeter. For configuration (a) we observe periodic variations of the spectral hole width, between $150$ and $450$~kHz, in synchrony with the rotary valve cycle.
For configurations (b) and (c) however, both datasets reveal very stable and narrow spectral holes, with no visible effect of the rotary valve cycle on the hole shape and width. The average hole width is $30$~kHz. This figure is remarkably low compared to the usual hole widths previously measured in Tm:YAG around 2-3~K with sub-kHz linewidth lasers and liquid helium cryostats (between $200$~kHz and $400$~kHz~\cite{deseze2006experimental,lauro2009slow}) and in Tm-doped crystals in general~\cite{macfarlane1993spectral,thiel2010optical,thiel2014ygg}. This result confirms the efficiency of the thermalization via the helium gas, together with the efficient vibration decoupling.
This also suggests that wet cryostats actually provide a not so quiet acoustic environment, susceptible to contribute to the spectral hole width.

\subsection{Vibration measurements}
\begin{figure*}[t]
\centering
\includegraphics[width=18cm]{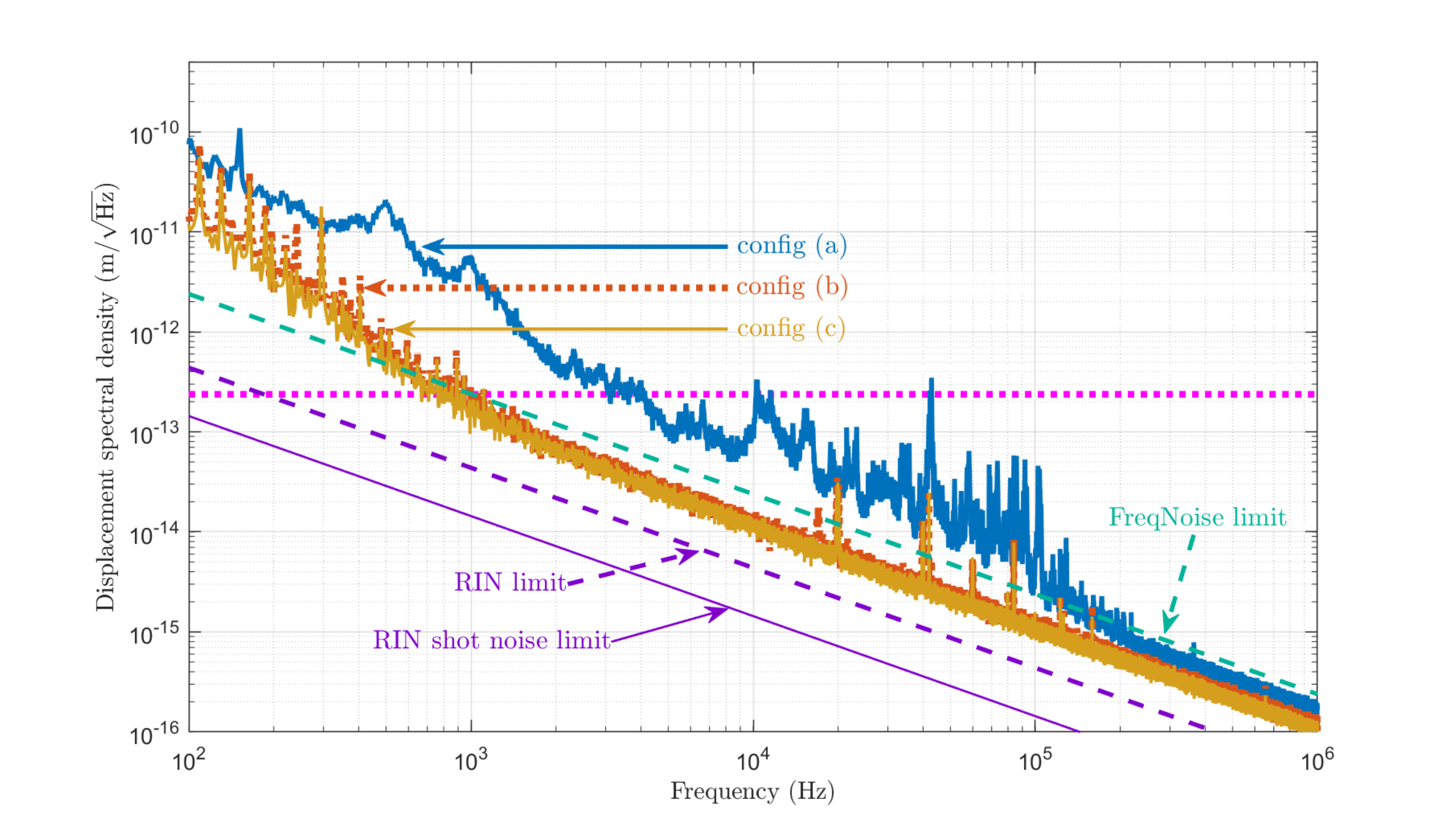}
\caption{Single-sided displacement spectral density obtained with our vibration sensor for the three configurations (a) (b) and (c) (see Table~\ref{table:exp}). Each spectrum represents the vibrations measured over three rotary valve cycles ($\simeq 1.5$~s). Dashed purple line: sensitivity limit due to the laser relative intensity noise (RIN). Solid purple line: sensitivity limit due to the shot noise limit of the RIN. Cyan dashed line: sensitivity limit due to the laser frequency noise.
Pink horizontal line: shot-noise limit that would apply for an interferometric measurement with a similar optical power.}
\label{fig:bruits}
\end{figure*}

\begin{figure}[ht]
\centering
\includegraphics[width=8.5cm]{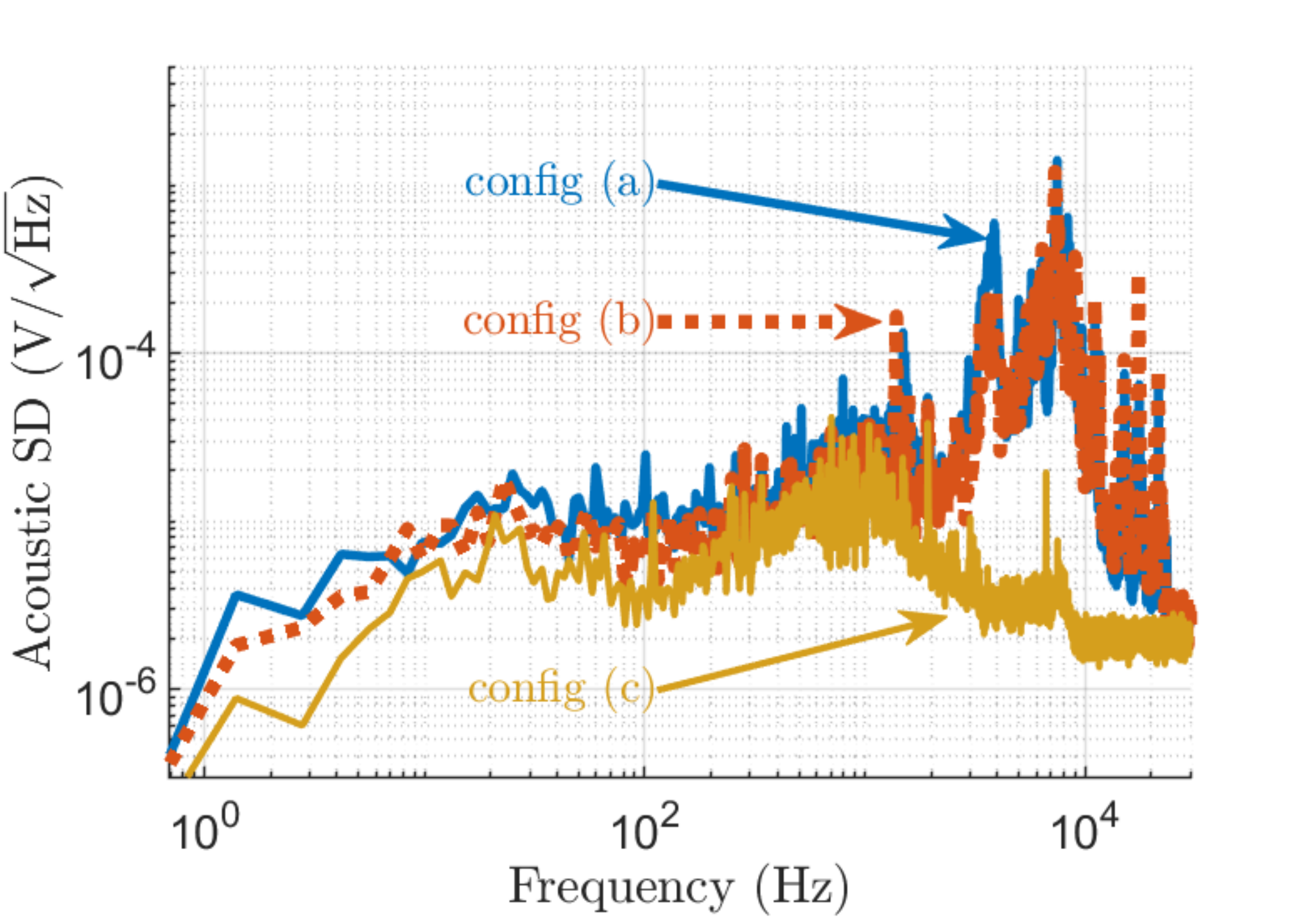}
\caption{Audio spectral density as measured by the soundmeter in contact with the rotary valve. In configurations (a) and (b), the sound sample reveals a dominating contribution of the cryocooler to the lab acoustic environment. In configuration (c) when the cryocooler is switched off, the remaining lab equipment can distinctly be heard, notably the turbomolecular vacuum pump (Pfeiffer HiCube80).}
\label{fig:acoustique}
\end{figure}

Now we apply our vibration measurement method to the three vibrational configurations (a) (b) and (c).
Based on the pump beam transmitted power over a duration $T$, we record the crystal absorption coefficient $\alpha(t)$ and we use Eq.~\ref{eq:abs} to derive the atomic velocity and infer the corresponding vibration-driven displacement.
The single-sided displacement spectral density is expressed in m\perroothz and is defined as:
\begin{equation}
A_x(f)=\sqrt{\frac 2 T}|\tilde{x}(f)|
\label{eq:Ax1}
\end{equation}
where $\sim$ denotes the Fourier transform. Equation~\ref{eq:Ax1} can also be written as a function of the atomic velocity, the relevant observable in our case (see Eq.~\ref{eq:abs}):
\begin{equation}
A_x(f)=\sqrt{\frac 2 T}\frac{1}{2\pi f}|\tilde{\dot{x}}(f)|
\label{eq:Ax2}
\end{equation}
We carry out a quasi-continuous acquisition of the crystal transmission over a $1.5$~s time interval (corresponding to $3$ rotary valve cycles). The data is acquired with a sampling rate of $2$ million samples per second. The acquisition is interrupted $1$~ms every $30$~ms to measure the spectral hole width $\Gamma_{\mathrm{HB}}$. On each $30$~ms interval, we convert the crystal absorption $\alpha(t)$ into the atomic velocity $\dot{x}(t)$ following Equation~\ref{eq:abs}, using the corresponding value of $\Gamma_{\mathrm{HB}}$.
Despite the periodic interruption, the high acquisition duty cycle (97\%) allows us to concatenate the data and derive the displacement spectral density between  $100$~Hz and $1$~MHz. The lower frequency limit is given by the inverse of the spectral hole lifetime ($T_1\simeq 10$~ms in Tm:YAG), while the higher frequency is fixed by the Shannon limit. We do this for the three vibrational configurations and present the results in Figure~\ref{fig:bruits}. We also plot the corresponding acoustic signal spectral density as measured by the soundmeter in Figure~\ref{fig:acoustique}.

In configuration (a) the vibration spectrum exhibits an overall behaviour close to $1/f$. A large number of peaks is observed between $150$~Hz and $200$~kHz that we attribute to the vibrations of the rotary valve transmitted to the crystal via the rigid assembly.

In configuration (b), the vibration spectrum drops significantly over a broad frequency range from $100$~Hz to $200$~kHz, although the audio spectrum is unchanged, indicating an efficient vibration decoupling provided by the suspension. Only in the $100-400$~Hz range does one observe an increase in the displacement spectral density in configurations (b) and (c), in the form of a series of peaks. We interpret these peaks as the result of a filtering of the acoustic noise by the $2$~cm-long suspension wires whose eigenfrequency is of the order of $100$~Hz.

\begin{figure}[t]
\centering
\includegraphics[width=8.7cm]{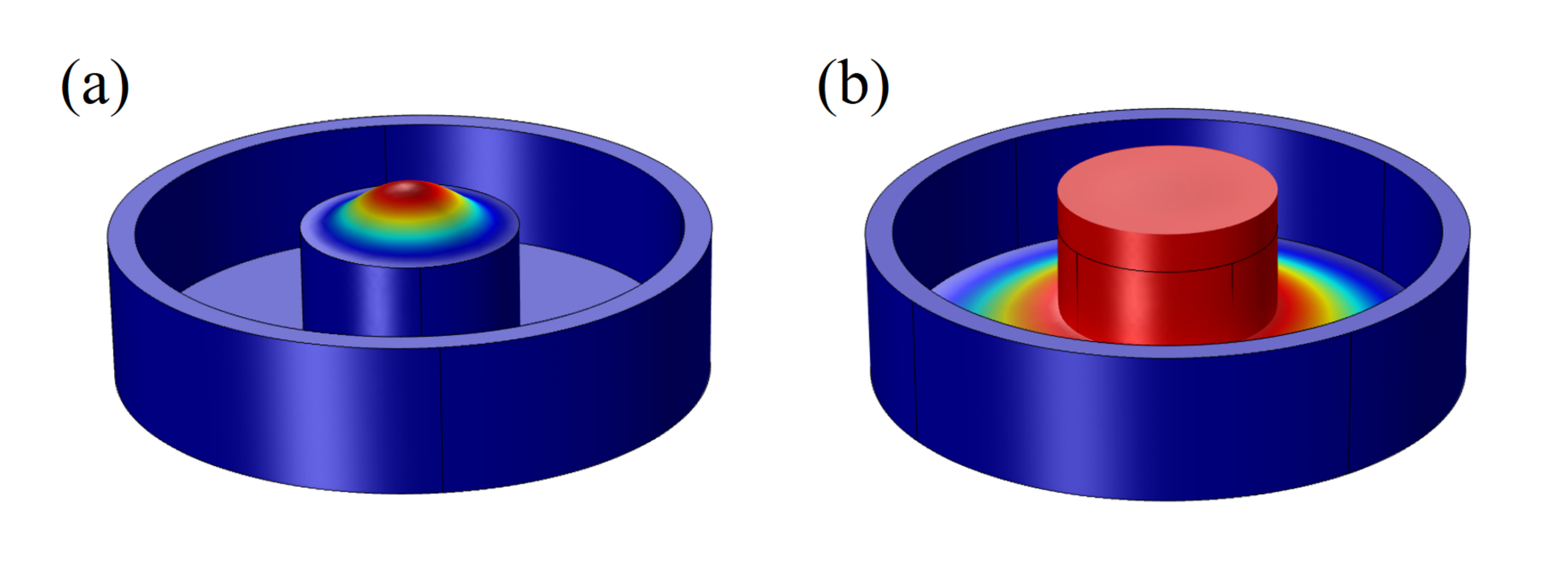}
\caption{Lowest order vibration modes for the sample holder. A COMSOL Multiphysics simulation gives an eigenfrequency of $70$~kHz and $7$~kHz for the $(0,1)$ drum mode for the center (a) and outer (b) platform.}
\label{fig:drum}
\end{figure}

The vibration spectrum in configuration (c) is identical to that of configuration (b), although the acoustic environment in the lab is much quieter especially in the $2-20$~kHz (see Figure~\ref{fig:acoustique}). This acoustic environment still contains significant contributions from other lab devices nearby that may propagate to the crystal via the helium gas or the suspension wires. Specifically we observe a few remaining peaks at $17$~kHz, $20$~kHz, $42$~kHz and $122$~kHz and their harmonics in the vibration spectra (b) and (c). These frequencies are compatible with the eigenfrequencies of low-order drum modes of the sample holder circular center platform (see Figure~\ref{fig:drum}). Besides, these peaks are also present in the vibration spectrum obtained in configuration (a). This is why we interpret them as due to the residual vibrations (non cryocooler-related), filtered by the sample holder mechanical resonances.

The invariance of the atomic displacement spectral density with respect to the operation of the compressor shows that we have reached the background sensitivity for our vibration sensing method. We will now investigate the various mechanisms that contribute to this background.

\subsection{Limits to the sensitivity}
In order to make apparent the various sources of noise, we linearize Equation~\ref{eq:abs}, assuming that the spectral hole shape is only marginally affected by the vibrations ($\alpha(t)\ll \alpha_0$):
\begin{equation}
|\dot{x}(t)|=\frac{\mathcal{V}}{\kappa E} \Gamma_{\mathrm{HB}}\frac{\alpha(t)}{\alpha_0}
\label{eq:xdot}
\end{equation}
The absorption coefficient is derived from the crystal transmission by using the Bouguer-Beer-Lambert absorption law linking the transmitted photon rate $I(t)$ to the incoming photon rate $I_0$: $I(t)=I_0 e^{-\alpha(t) L}$. Based on the fact that $\alpha_0L$ is of the order of unity, we obtain:
\begin{equation}
|\dot{x}(t)|=\frac{\mathcal{V}}{\kappa E} \frac{\Gamma_{\mathrm{HB}}}{\alpha_0 L}\left( 1-\frac{I(t)}{I_0}\right)
\end{equation}
The displacement spectral density can therefore be written as:
\begin{equation}
A_x(f)=\frac{1}{2\pi f}\frac{\mathcal{V}}{\kappa E} \frac{\Gamma_{\mathrm{HB}}}{\alpha_0 L}\underbrace{\frac{A_I(f)}{I_0}}_{\sqrt{\mathrm{RIN}}}
\label{eq:Axf}
\end{equation}
In this expression the parameters $\Gamma_{\mathrm{HB}}$ and $\alpha_0L$ are measured experimentally. The ratio $\frac{A_I(f)}{I_0}$ corresponds to the square root of the transmitted laser relative intensity noise (RIN).

We estimate the incoming relative intensity noise of our ECDL laser itself by an independent measurement with a similar power: $ \left. \frac{A_I(f)}{I_0}\right\vert_{\mathrm{RIN}}=2.5\times 10^{-6}$\perroothz.
Using Equation~\ref{eq:Axf} we derive an estimation of the RIN-limited sensitivity floor for our displacement measurement, taking $\alpha_0L=1$ and $\Gamma_{\mathrm{HB}}=300~$kHz, a typical spectral hole width measured in configuration (a) during the noisy phase of the rotary valve cycle [see Figure~\ref{fig:trous}(a)]:
\begin{equation}
A_x^{\mathrm{RIN}}(f)= \frac{4.4}{f} \times 10^{-11}~\textrm{m\perroothz}
\label{eq:AxRIN}
\end{equation}

The laser RIN is a technical issue, ultimately limited by the laser intensity shot noise $\left. \frac{A_I(f)}{I_0}\right\vert_{\mathrm{shot}}=\sqrt{\frac 2 {I_0}}$ where $I_0$ is the photon rate. Considering the typical laser power collected on the detector in the present work ($0.7~\mu$W), this contribution amounts to $8.3 \times 10^{-7}$\perroothz.
Again, using Equation~\ref{eq:Axf} we obtain a fundamental limit for the piezospectroscopic vibration measurement:
\begin{equation}
A_x^{\mathrm{shot}}(f)= \frac{1.4}{f} \times 10^{-11}~\textrm{m\perroothz}
\label{eq:Axshot}
\end{equation}
One should bear in mind that this shot noise limit is affected by a $\Gamma_{\mathrm{HB}}/\alpha_0L$ scaling factor (see Eq.~\ref{eq:Axf}). While the spectral hole width $\Gamma_{\mathrm{HB}}$ is ultimately limited by twice the homogeneous linewidth ($\Gamma_{\mathrm{HB}}\geq 2\Gamma_h=6~$kHz in Tm:YAG~\cite{macfarlane1993photon}), reaching such a low limit would require both a very low vibration level and a reduced laser irradiance on the crystal to avoid optical saturation, in turn leading to a reduced hole depth $\alpha_0$.


%

Another noise source that can contribute to the vibration background is the laser frequency noise. Considering that the pump beam mean frequency is by definition centered on the spectral hole, we approximate the lorentzian hole shape at the lowest order:
\begin{equation}
\alpha(\delta(t))=4 \alpha_0 \frac{\delta(t)^2}{\Gamma_{\mathrm{HB}}^2}
\end{equation}
where $\delta(t)$ is the time-dependent  detuning between the laser frequency and the center of the spectral hole.
Going back to Equation~\ref{eq:xdot}, it appears that the laser frequency noise only contributes to second order to the atomic velocity:
\begin{equation}
\left|\dot{x}(t)\right| = 4 \frac{\mathcal{V}}{\kappa E} \frac{1}{\Gamma_{\mathrm{HB}}} \delta(t)^2
\end{equation}
If $\delta(t)$ is a centered white noise, one can show that $A_{\delta^2}(f)= 2 A_{\delta}(f)^2/\sqrt{\Delta t}$, where $\Delta t$ is the sampling period (see appendix). Therefore the displacement spectral density originating from the laser frequency noise reads as:
\begin{equation}
A_x^{\mathrm{FreqN}}(f)= \frac{4}{\pi f}\frac{\mathcal{V}}{\kappa E} \frac{1}{\Gamma_{\mathrm{HB}}\sqrt{\Delta t}} A_{\delta}(f)^2
\end{equation}
Interestingly, the contribution of the laser frequency noise to the vibration measurement sensitivity is inversely proportional to the spectral hole width. Taking $\Gamma_{\mathrm{HB}}=300$~kHz, $\Delta  t=0.5~\mu$s and $A_{\delta}(f)=10$~Hz\perroothz \ (according to previous measurements made on our laser source~\cite{crozatier2004laser}), we obtain the following estimation for the laser frequency-noise contribution to the displacement spectral density:
\begin{equation}
A_x^{\mathrm{FreqN}}(f)= \frac{24}{f} \times 10^{-11}~\textrm{m\perroothz}
\label{eq:AxFN}
\end{equation}

We plot the noise contribution estimations given in equations~\ref{eq:AxRIN}, \ref{eq:Axshot} and \ref{eq:AxFN} in Figure 3 together with the experimental data. Among these three possible contributions to the noise floor of our vibration sensor, the laser frequency noise appears to be the most significant one and is compatible with the experimental observation. It should be noted that in order to perform our estimation, we assume a white noise for the laser frequency that may differ from the real noise spectrum especially at high frequencies.

Finally, we observe that the contributions of frequency noise and intensity noise depend on the spectral hole width $\Gamma_{\mathrm{HB}}$ in opposite ways: a narrower spectral hole would raise the frequency noise contribution and lower the intensity noise contribution. It is interesting to note that even with spectral holes of several tens of kHz, a sub-kHz linewidth laser still exhibits excessive frequency noise limiting the sensitivity of our device.
This study highlights the importance of working with a laser source exhibiting minimal frequency noise, especially when dealing with narrow spectral holes.

Overall, we have identified the sensitivity limits of our vibration sensor. Even in the quietest  configuration, the remaining acoustic perturbations transmitted to the crystal lead to a series of peaks that limit the sensitivity below $1$~kHz. At higher frequencies, the sensor sensitivity reaches a $1/f$ floor limited by the laser frequency noise.

\section{Discussion}
Our piezospectroscopic vibration sensor fundamentally differs from conventional optical methods (\emph{ie}, interferometric) because its sensitivity is based on strain (and not on position). This is why the sensitivity increases with the  frequency. In this work we have reached $1$~MHz but this is a mere technical limitation given by the $2$~MHz sampling rate. More specifically, we demonstrate a $10^{-16}$~m\perroothz background displacement sensitivity at $1$~MHz with a $0.7~\mu$W optical power, whereas its counterpart in an interferometric measurement would only reach $\frac{\lambda}{2\sqrt{I_0}}=2.4\times 10^{-13}$~m\perroothz with the same power. This interferometric measurement sensitivity is displayed in Figure~\ref{fig:bruits}. Reaching the sensitivity of our piezospectroscopic method at $1$~MHz with an interferometric setup would require working with an interferometer with a high finesse ($>1000$).

Besides sensitivity, we point out that our method allows for inertial sensing, unlike conventional optical methods that give access to the displacement with respect to a reference point, which could itself be unstable. This property is also a direct consequence of the strain-coupling mechanism of our sensor. This is a decisive asset of our method for the most vibration-sensitive setups, including those using micro or nano-mechanical resonators, where inertial stability is required.

\section{Conclusion}
In a previous publication~\cite{louchet2019piezospectroscopic} we have proposed an original inertial method to probe acoustic vibrations in a cryogenic environment on a wide frequency range, based on spectral holeburning in a rare-earth ion-doped crystal.
In the present work we have investigated the technical and fundamental limits of this method. This was made possible by exploring various vibrational configurations in a pulse-tube cryocooler. In the quietest environment combining mechanical decoupling from the cold finger and momentary pause of cryocooler operation, we have observed the narrowest spectral holes ever reported in a Tm:YAG crystal. Above $1$~kHz we have reached a $1/f$ background floor that we ascribe to laser frequency noise.  Below $1$~kHz, the excess of noise is attributed to the remainder of the strongly attenuated mechanical vibrations transmitted to the crystal by the suspension and the helium gas, characterizing the limit of our vibration-decoupling setup.

\section{Acknowledgments}
The authors acknowledge support from the French National Research Agency (ANR) projects ATRAP (ANR-19-CE24-0008),
and MIRESPIN (ANR-19-CE47-0011), and the LABEX
WIFI (Laboratory of Excellence within the French Program "Investments for the Future") under references
ANR-10-LABX-24 and ANR-10-IDEX-0001-02 PSL*. 

\section{Authors declarations}
The authors have no conflicts of interest to disclose.

\section{Data availability}
The data that support the findings of this study are available from the corresponding author upon reasonable request.

\appendix
\section{White noise properties}
In this appendix we establish a relationship between the two-sided power spectral density (PSD) of a white noise $x$ and the PSD of its square $x^2$.

Let $x$ be a centered white noise with variance $\sigma_x^2$, sampled with a period $\Delta t$. Its PSD is constant over the whole measurement bandwidth:
\begin{equation}
    \textrm{if }|f|<\frac{1}{2\Delta t},  \ S_x(f)=S_x^0=\sigma_x^2 \Delta t
\label{eq:sxf}
\end{equation}

The quantity $x^2$ is not strictly a white noise since it has a non-zero mean value, so one cannot directly write an equivalent property for $x^2$. However, the random process $u=x^2-E(x^2)$ is a centered white noise and we can write equation~\ref{eq:sxf} for $u$:
\begin{equation}
    S_u^0=\sigma_u^2 \Delta t
\end{equation}
Noting that $u$ and $x^2$ have the same PSD for all frequencies in the bandwidth except for $f=0$, we obtain:
\begin{equation}
    \textrm{if } 0<|f|<\frac{1}{2\Delta t},\  S_{x^2}(f)=S_u^0=\sigma_u^2 \Delta t
\end{equation}
Keeping in mind that $\sigma_u^2=2 \sigma_x^4$, and using \ref{eq:sxf} we finally get:
\begin{equation}
    \textrm{if } 0<|f|<\frac{1}{2\Delta t},\ S_{x^2}(f)=\frac{2}{\Delta t} S_x(f)^2
\end{equation}

We therefore obtain the following relationship for single-sided amplitude spectral densities defined as $A_x(f)=\sqrt{2S_x(f)}$:
\begin{equation}
    A_{\delta^2}(f)= 2 A_{\delta}(f)^2/\sqrt{\Delta t}
\end{equation}

%

\end{document}